\documentclass[a4paper,11pt]{article}
\usepackage[utf8]{inputenc}
\usepackage[english]{babel}
\usepackage{braket}
\usepackage{xspace}
\usepackage{bbm}
\usepackage{mathdots}
\usepackage{stackrel}
\usepackage[dvipsnames]{xcolor}
\usepackage{appendix}
\usepackage{hyperref}
\hypersetup{
 colorlinks=true,
 linkcolor=blue,
 anchorcolor = blue,
 citecolor = blue,
 filecolor = blue,
 urlcolor = blue
}
\usepackage{mathtools}
\usepackage{bbm}
\usepackage{comment}
\usepackage{subfigure}
\usepackage[T1]{fontenc}               
\usepackage[left=3cm,
            right=2.5cm,
            top=2.5cm,
            bottom=3cm
            ]{geometry}                
\usepackage{graphicx}
\usepackage{mathrsfs}
\usepackage{appendix}
\usepackage{fancyhdr}
\usepackage{amsmath,                   
            amssymb,                   
            amsthm}                    

\usepackage{setspace}
\usepackage{cite}
\usepackage{cancel}
\usepackage{bm}
\usepackage[margin=30pt, bf, font=small, center, justification=justified]{caption}[2004/07/16]

\usepackage{setspace} 

\usepackage{tikz}
\usepackage{bbm}

\def\Tr{{\rm Tr}}

\title{\bf  Entanglement asymmetry and quantum Mpemba effect for Kramers-Wannier duality
}

\author{Milo Vescovo$^{1,2}$, Pasquale Calabrese$^1$, Filiberto Ares$^1$}

\date{}

\begin{document} 

\maketitle

{\small
\vspace{-5mm}  \ \\
{$^{1}$}  SISSA and INFN, via Bonomea 265, 34136 Trieste, Italy\\[-0.1cm]
\medskip
{$^{2}$}  ENS Paris-Saclay, France
}

\begin{abstract}

The Kramers-Wannier duality is the prototypical example of a non-invertible symmetry, yet little is known about its fate away from criticality and out of equilibrium. We introduce the \emph{Kramers-Wannier entanglement asymmetry}, a quantum-information measure that quantifies the breaking of this non-invertible symmetry in the transverse-field Ising chain. We first investigate its equilibrium properties, showing that it exhibits a striking crossover between the ordered and disordered phases together with a pronounced dip at the critical point that becomes increasingly sharp with subsystem size. We then study quantum quenches from both gapped phases to criticality and show that the Kramers-Wannier entanglement asymmetry decays to zero, signaling the dynamical restoration of the non-invertible symmetry. Remarkably, we uncover the emergence of a quantum Mpemba effect: under suitable conditions, states initially farther from equilibrium restore the Kramers-Wannier symmetry faster than states prepared closer to it. We provide both analytical and numerical evidence for this phenomenon and identify the mechanism responsible for its occurrence. Our work establishes entanglement asymmetry as a powerful probe of non-invertible symmetries beyond equilibrium and opens new perspectives on the dynamics of dualities in quantum many-body systems.

\end{abstract}
\newpage

\tableofcontents
\section{Introduction}

The Kramers-Wannier duality~\cite{kw-41} is the oldest and arguably the best-known example of a non-invertible symmetry, a concept that has recently emerged as a unifying theme across condensed matter physics, quantum field theory, and high-energy physics~\cite{gksw-15,amf-16,cdis-22,s-24,sn-24, bt-18, afm-20}. Originally introduced as a remarkable duality relating the low- and high-temperature phases of the two-dimensional classical Ising model, it has profoundly shaped our understanding of phase transitions and critical phenomena. Among its many celebrated applications, the Kramers-Wannier duality provides an elegant and exact derivation of the critical point of the two-dimensional classical Ising model (or, equivalently, of the one-dimensional quantum Ising chain) as the self-dual point. More generally, it has become the prototype for understanding non-invertible symmetries and the algebraic structures underlying them. Despite decades of study, however, the Kramers-Wannier duality continues to reveal new facets and to challenge our understanding of non-invertible symmetries.

One aspect that has received comparatively little attention concerns the fate of the Kramers-Wannier duality away from criticality (some results are only present in field theory~\cite{benini-25,bgvv-26,ggh-25}) and, even more so, out of thermal equilibrium. While the role of non-invertible symmetries at critical points is by now rather well understood, much less is known about how these symmetries are broken and subsequently restored in generic nonequilibrium settings. Addressing this question is the main goal of the present work.

To this end, we generalize the recently introduced concept of \emph{entanglement asymmetry}~\cite{amc-23} (see also \cite{mhms-22,vaw-08,chmp-20,chmp-21,ms-14}), originally devised to quantify the breaking of ordinary (invertible) symmetries, to the case of the Kramers-Wannier duality in the transverse-field quantum Ising chain. This construction provides a quantitative information-theoretic measure of the breaking of a non-invertible symmetry and allows us to investigate its behavior both in equilibrium and during unitary time evolution.
We consider two physically distinct and complementary situations. First, we analyze the ground state of the quantum Ising chain away from criticality. We show that the Kramers-Wannier entanglement asymmetry exhibits a striking crossover between the ferromagnetic and paramagnetic phases, together with a pronounced dip at the critical point that becomes increasingly sharp as the subsystem size grows.

We then turn to nonequilibrium dynamics and investigate quantum quenches from both gapped phases to the critical point. We characterize how the non-invertible symmetry is dynamically restored, demonstrating that the corresponding entanglement asymmetry vanishes asymptotically at long times. Remarkably, we also uncover the emergence of a quantum Mpemba effect~\cite{amc-23,acm-25,c-26,tbl-26,joshi-24} for the Kramers-Wannier entanglement asymmetry: under suitable conditions, states prepared further away from equilibrium restore the non-invertible symmetry faster than states that are initially closer to the critical point. Combining analytical arguments with numerical simulations, we identify the mechanism responsible for this phenomenon and establish precise conditions under which the Mpemba effect occurs.

Overall, our results provide a first systematic characterization of the nonequilibrium dynamics of a non-invertible symmetry in lattice models through the lens of quantum information, opening a new direction in the study of dualities and their dynamical restoration beyond criticality.

The remainder of this manuscript is organized as follows. In Sec.~\ref{sec2}, we introduce the Kramers-Wannier (KW) entanglement asymmetry and evaluate it for several simple states. In Sec.~\ref{sec3}, we review the transverse-field Ising chain, describe the KW duality in the Majorana representation, and summarize the correlation-matrix formalism used to compute entanglement-related quantities. In Sec.~\ref{sec4}, we investigate the equilibrium properties of the KW entanglement asymmetry, providing both analytical and numerical results for the ground state away from the critical point. In Sec.~\ref{sec5}, we study the nonequilibrium dynamics of the KW entanglement asymmetry following a quench to the critical point and develop a quasiparticle description of its time evolution. Finally, in Sec.~\ref{sec6}, we summarize our results and discuss several open problems. Technical details and complementary calculations are collected in the appendices.

\section{Kramers-Wannier duality and entanglement asymmetry}
\label{sec2}

In a spin-$1/2$ chain, the Kramers-Wannier duality maps the Pauli matrices $\sigma_j^\mu$, $\mu=x, y, z$, at site $j$ to the dual Pauli matrices~\cite{kw-41, radicevic-18}
\begin{equation}\label{eq:kw-spins}
\hat{\sigma}_j^z=\sigma_j^x\sigma_{j+1}^x,\quad \hat{\sigma}_j^x=\prod_{l\leq j}\sigma_l^z.
\end{equation}
The dual Pauli matrices $\hat{\sigma}_j^\mu$ satisfy the same Pauli algebra as the original ones,
although they are non-local when expressed in terms of the original spins. We note that on a finite chain the implementation of the Kramers–Wannier duality requires care with the boundary conditions. For instance, with periodic boundary conditions, the different $\mathbb{Z}_2$ parity sectors of the original Hilbert space are mapped to different boundary conditions (periodic or antiperiodic) for the dual spins. To avoid these subtleties, we always work with infinite spin chains.

Let us focus on a spin chain in a state described by the density matrix $\rho$. The Kramers-Wannier dual state is obtained by mapping operators according to the duality transformation~\eqref{eq:kw-spins}, i.e. by defining a new density matrix $\hat{\rho}$ 
such that expectation values of dual observables are preserved,
\begin{equation}\label{eq:kw_tr}
{\rm Tr}(\rho O)={\rm Tr}(\hat{\rho}\hat{O}),
\end{equation}
where $\hat{O}$ is obtained from $O$ by replacing $\sigma_j^\mu \mapsto \hat{\sigma}_j^\mu$ 
using the relations~\eqref{eq:kw-spins}. The state $\rho$ is self-dual if it remains invariant under Eq.~\eqref{eq:kw-spins}, i.e. if $\hat{\rho} = \rho$.

The entanglement asymmetry quantifies the extent to a quantum state $\rho$ breaks a given symmetry~\cite{amc-23}. 
It is defined as the relative entropy between the state and its symmetrization,
\begin{equation}
\Delta S = S(\rho||\rho_S) = -{\rm Tr}\left[\rho(\log\rho - \log\rho_S)\right],
\end{equation}
where $\rho_S$ is the state obtained by symmetrizing $\rho$ with respect to the symmetry under consideration.  
In general, the entanglement asymmetry is a non-negative observable, $\Delta S\geq0$, which vanishes if an only if $\rho$ is symmetric, i.e. $\rho_S=\rho$. This definition was recently extended in Ref.~\cite{benini-25} to higher-form and non-invertible symmetries, including the Kramers–Wannier duality.

In the case of the Kramers–Wannier duality, the symmetrized density matrix is obtained by averaging over the dual transformation,
\begin{equation}\label{eq:symmetrization}
\rho_S=\frac{\rho+\hat{\rho}}{2},
\end{equation}
where $\hat{\rho}$ is the Kramers–Wannier dual of $\rho$. By construction, $\rho_S$ is self-dual, i.e. $\hat{\rho}_S=\rho_S$. 
By inserting  the symmetrized state into the relative entropy definition and using the properties of the Kramers–Wannier duality, in particular Eq.~\eqref{eq:kw_tr}, 
the entanglement asymmetry can be rewritten as the difference of von Neumann entropies,
\begin{equation}\label{eq:EA_diff}
\Delta S = S(\rho_S) - S(\rho),
\end{equation}
where $S(\rho)=-{\rm Tr}(\rho\log\rho)$. 
In general, the calculation of the entanglement entropies in Eq.~\eqref{eq:EA_diff} is challenging. 
A way to bypass this problem is to do the replica trick, by replacing in that equation the von Neumann entropy by the R\'enyi
entropy $S^{(n)}(\rho)=1/(1-n)\log{\rm Tr}(\rho^n)$. We therefore introduce the R\'enyi-$n$ KW asymmetry~\cite{amc-23},
\begin{equation}\label{eq:renyi_EA}
\Delta S^{(n)}=S^{(n)}(\rho_S) - S^{(n)}(\rho).
\end{equation}
For integer values of $n$, the Rényi entanglement asymmetry is generally much easier to compute than its von Neumann counterpart. 
The entanglement asymmetry in Eq.~\eqref{eq:EA_diff} is then recovered by analytically continuing $\Delta S^{(n)}$ to real values of $n$ and taking the limit $n\to1$.
Moreover, the Rényi entanglement asymmetry is experimentally accessible via randomized measurements and classical shadows techniques, and 
has been recently measured in various experimental settings~\cite{joshi-24, yang-26, xu-26}.

As a first illustration of the KW  asymmetry, we consider the following states of an infinite spin-$1/2$ chain,
\begin{equation}\label{eq:pm_f}
\ket{+}=\ket{\rightarrow\rightarrow\dots\rightarrow},\quad
\ket{-}=\ket{\leftarrow\leftarrow\dots\leftarrow}, \quad
\ket{f}=\ket{\uparrow \uparrow\dots \uparrow}.
\end{equation}
Here $\ket{\uparrow}$ and $\ket{\downarrow}$ are the eigenstates of $\sigma^z$, while $\ket{\rightarrow}$ and $\ket{\leftarrow}$ are the eigenstates of $\sigma^x$.
In the thermodynamic limit, the states in Eq.~\eqref{eq:pm_f} are orthonormal.
Under the Kramers–Wannier duality~\eqref{eq:kw-spins}, they transform as
\begin{equation}\label{eq:kw_f_pm}
\ket{f}\mapsto \frac{\ket{+}+\ket{-}}{\sqrt{2}}, \quad
\ket{\pm}\mapsto \ket{f}.
\end{equation} 
If we consider $\rho=\ket{\pm}\bra{\pm}$, its symmetrized density matrix is
\begin{equation}
\rho_S=\frac{\ket{\pm}\bra{\pm}+\ket{f}\bra{f}}{2},
\end{equation}
and its Rényi-$n$ KW asymmetry is in the thermodynamic limit
\begin{equation}\label{eq:EA_p}
\Delta S^{(n)}(\ket{\pm})=\log 2.
\end{equation}
If we instead take the state $\rho=\ket{f}\bra{f}$, then the symmetrized density matrix reads
\begin{equation}
\rho_S=\frac{1}{2}\ket{f}\bra{f}+\frac{1}{4}(\ket{+}+\ket{-})(\bra{+}+\bra{-}).
\end{equation}
and the entanglement asymmetry is again
\begin{equation}\label{eq:EA_f}
\Delta S^{(n)}(\ket{f})=\log 2
\end{equation}
in the thermodynamic limit.
In the critical Ising chain, the states in Eq.~\eqref{eq:pm_f} correspond in the scaling limit to conformal boundary states of the Ising CFT: $\ket{f}$ flows to the free boundary condition, while $\ket{\pm}$ flow to the two fixed boundary conditions~\cite{cardy-86,cardy-89, konechny-17, arv-20, fi-21}. The KW  asymmetry for these boundary states has been 
investigated in Ref.~\cite{benini-25}, obtaining the same result as in Eqs.~\eqref{eq:EA_p} and~\eqref{eq:EA_f}. 

We can also consider the KW  asymmetry in a subsystem $A$ consisting of $\ell$ contiguous sites. 
If the full system is in a state $\rho$, the subsystem is described by the reduced density matrix $\rho_A = \mathrm{Tr}_{\bar A}\rho$, 
obtained by tracing out the complement $\bar A$. However, in this case the Kramers–Wannier duality does not act within the algebra 
of operators supported on $A$, since it is a non-local transformation that mixes degrees of freedom inside and outside the subsystem, 
mapping local operators into non-local string operators that generally extend across the boundary of the subsystem.. 
As a consequence, a direct construction of the dual reduced density matrix from Eq.~\eqref{eq:kw-spins} is not possible in general.
One way to address this issue is to extend the subsystem by introducing suitable boundary degrees of freedom at its endpoints, so 
that the Kramers–Wannier duality can be consistently implemented at the level of the reduced density matrix.
In this work, instead, we take an alternative approach. We first apply the Kramers–Wannier duality to the full density matrix $\rho$, 
obtaining $\hat{\rho}$, and then define the dual reduced density matrix as $\hat{\rho}_A = \mathrm{Tr}_{\bar A}(\hat{\rho})$. 
This procedure avoids any ambiguity associated with attempting to define the duality directly on the reduced density matrix. 
 This construction was previously introduced for translational symmetry in Refs.~\cite{krb-25, ggsb-25} where similar problems occur. For internal invertible symmetries, the procedure is unambiguous, since symmetrization and taking the partial trace commute, so that performing these operations in either order leads to the same reduced density matrix. We will refer to the asymmetry obtained in this way as the {\it Kramers-Wannier entanglement asymmetry}.

If we apply this method to the state $\rho=\ket{f}\bra{f}$, the partial trace is trivial since it is a pure state, $\rho_A=\ket{f_A}\bra{f_A}$, 
where the subscript denotes the restriction of the product state $\ket{f}$ to the sites in $A$. On the other hand, the partial trace of its dual state 
$\hat{\rho}=(\ket{+}+\ket{-})(\bra{+}+\bra{-})/2$ is the mixed state
\begin{equation}
\hat{\rho}_A=\frac{\ket{+_A}\bra{+_A}+\ket{-_A}\bra{-_A}}{2}.
\end{equation}
Inserting $\rho_A$ and $\hat{\rho}_A$ in Eq.~\eqref{eq:symmetrization}, the symmetrized state reads 
\begin{equation}\label{eq:rho_AS_f}
\rho_{A,S}=\frac{1}{4}(\ket{+_A}\bra{+_A}+\ket{-_A}\bra{-_A})+\frac{1}{2}\ket{f_A}\bra{f_A},
\end{equation}
and the KW
entanglement asymmetry of the subsystem is 
\begin{equation}
\label{eq:EA_f_subs}
\Delta S_A^{(n)}(\ket{f})=\frac{1}{1-n}\log\left[4^{-n}+\left(\frac{3}{8}+\frac{\sqrt{1+2^{4-\ell}}}{8}\right)^n+\left(\frac{3}{8}-\frac{\sqrt{1+2^{4-\ell}}}{8}\right)^n\right].
\end{equation}
In the limit $\ell\to\infty$, it tends to the finite value
\begin{equation}
\Delta S_A^{(n)}(\ket{f})\to \frac{1}{1-n}\log\left(\frac{2^n+2}{2^{2n}}\right).
\label{DSf}
\end{equation}
Observe that, for the state $\ket{f}$, the KW asymmetry of a subsystem~\eqref{eq:EA_f_subs} is larger than the asymmetry of the total system~\eqref{eq:EA_p}, i.e. $\Delta S_A^{(n)}(\ket{f})\geq \log 2 $. This result follows from the fact that the dual of $\ket{f}$ is not a product state.
In Ref.~\cite{benini-25}, the same result~\eqref{DSf} was found for the asymmetry of the boundary states corresponding to $\ket{\pm}$ with respect to the full fusion algebra of the Ising CFT. In this case, the state is symmetrized simultaneously with respect to the $\mathbb{Z}_2$ symmetry and the KW duality, yielding the same symmetrized density matrix as in Eq.~\eqref{eq:rho_AS_f} and hence to the same KW asymmetry.

In contrast, when we consider the states $\ket{\pm}$, the Kramers-Wannier dual is a product state, $\hat{\rho}=\ket{f}\bra{f}$, and the KW asymmetry of a subsystem is 
\begin{equation}
\Delta S_A^{(n)}(\ket{\pm})=
\frac{1}{1-n}\log\left[\frac{(1+2^{-\ell/2})^n+(1-2^{-\ell/2})^n}{2^n}\right],
\end{equation}
which in the limit $\ell\to\infty$ coincides with the total-system result in Eq.~\eqref{eq:EA_p}. 

Finally, the KW entanglement asymmetry of the linear combination $(\ket{+}+\ket{-})/\sqrt{2}$ will also be relevant for the discussion below. Under the transformations~\eqref{eq:kw_f_pm}, its normalized KW dual state is $\ket{f}$. Therefore, the symmetrized reduced density matrix coincides with the one obtained in Eq.~\eqref{eq:rho_AS_f}. However, the two cases differ in the reduced density matrix of the original state: while $\ket{f}$ is a product state, and the subsystem entropy is zero, the state $(\ket{+}+\ket{-})/\sqrt{2}$ has non-trivial entanglement. As a result, the corresponding KW entanglement asymmetries are different. For the state $(\ket{+}+\ket{-})/\sqrt{2}$, we have (cf. Eqs.~\eqref{eq:EA_f_subs}-\eqref{DSf})
\begin{equation}
\label{eq:EA_lc_pm}
\Delta S_A^{(n)}\left(\frac{\ket{+}+\ket{-}}{\sqrt{2}}\right)=\frac{1}{1-n}\log\left[2^{-1-n}+\frac{\left(3+\sqrt{1+2^{4-\ell}}\right)^n+\left(3-\sqrt{1+2^{4-\ell}}\right)^n}{2^{1+2n}}\right],
\end{equation}
which in the limit $\ell\to\infty$ gives
\begin{equation}
\Delta S_A^{(n)}\left(\frac{\ket{+}+\ket{-}}{\sqrt{2}}\right)\rightarrow \frac{1}{1-n}\log\left(\frac{2^n+2}{2^{n+1}}\right).
\label{DS+-}
\end{equation}

In the subsequent sections, we extend these results to the ground state of a paradigmatic spin-$1/2$ chain: the transverse-field Ising model.

\section{Kramers-Wannier in the Ising chain in Majorana language}
\label{sec3}

In this section, we review the KW duality in  the transverse-field Ising spin-$1/2$ chain and introduce the basic tools to study the KW entanglement asymmetry. The Hamiltonian of this system is \cite{s-book}
\begin{equation}\label{eq:ising_hamiltonian}
		H(h) = -\frac{1}{2}\sum_{j \in \mathbb{Z}} [\sigma^{x}_j \sigma^{x}_{j+1} +h\sigma^z_j],
\end{equation}
where $h\geq 0$ is a transverse external magnetic field. As well-known, this model exhibits two phases: a ferromagnetic (ordered) phase for $0\leq h<1$ and a paramagnetic (disordered) phase for $h>1$. These are separated by the critical point $h=1$, where the system is gapless and its low-energy behavior is described by the Ising CFT. 
Under the KW transformation~\eqref{eq:kw-spins},  the Hamiltonian~\eqref{eq:ising_hamiltonian} transforms as 
$H(h)\mapsto \hat{H}(h)=h H(1/h)$. The Kramers–Wannier duality thus relates the ferromagnetic and paramagnetic phases by mapping $h\leftrightarrow 1/h$. Consequently, the critical point $h=1$, which is left invariant under the duality, is identified as the self-dual point of the model.

To study the KW entanglement asymmetry in the Ising chain, it is useful to map it to a chain of spinless fermions via the Jordan–Wigner transformation,
\begin{equation}\label{eq:jw_original}
\gamma_{2j}=\left(\prod_{l = 1}^{j-1}\sigma^{z}_l\right)\sigma_j^{x}, \quad \gamma_{2j-1}=\left(\prod_{l = 1}^{j-1}\sigma^{z}_l\right)\sigma_j^{y},
\end{equation}
where $\gamma_j$ are Majorana fermionic operators that satisfy the anticommutation relations $\{\gamma_j, \gamma_{j'}\}=2\delta_{jj'}$. In terms of them, the Ising Hamiltonian~\eqref{eq:ising_hamiltonian} is the quadratic form
\begin{equation}\label{eq:majorana_Ham}
H(h) = \frac{i}{2}\sum_{j \in\mathbb{Z}} [\gamma_{2j}\gamma_{2j+1} +h\gamma_{2j-1}\gamma_{2j}].
\end{equation}
Using the Jordan-Wigner transformation, we can also derive a fermionic representation of the KW duality. Let $\hat{\gamma}_j$ 
denote the Majorana operators obtained by applying the Jordan-Wigner transformation to the dual Pauli matrices $\hat{\sigma}_j^\mu$. They are defined as
\begin{equation}
\hat{\gamma}_{2j}=\left(\prod_{l = 1}^{j-1}\hat{\sigma}^{z}_l\right)\hat{\sigma}_j^{x}, \quad \hat{\gamma}_{2j-1}=\left(\prod_{l = 1}^{j-1}\hat{\sigma}^{z}_l\right)\hat{\sigma}_j^{y},
\end{equation}
and therefore satisfy $\{\hat{\gamma}_j, \hat{\gamma}_{j'}\}=2\delta_{jj'}$. Combining the relations between the original and dual Pauli operators in Eq.~\eqref{eq:kw-spins} and the Jordan-Wigner transformation for the latter~\eqref{eq:jw_original}, we find the following transformation of the Majorana operators:
\begin{equation}\label{eq:KW_majorana}
		\hat{\gamma}_{2n-1} = -i \gamma_1 \gamma_{2n}, \quad \hat{\gamma}_{2n} = -i\gamma_1 \gamma_{2n+1}.
\end{equation}
Thus, in terms of Majorana fermions, the KW duality takes the form of a translation by one site up to the additional factor $-i\gamma_1$~\cite{ss-24, sss-24}. Correspondingly, the dual Hamiltonian,
\begin{equation}
\hat{H}(h) = \frac{i}{2}\sum_{j \in\mathbb{Z}} [\hat{\gamma}_{2j}\hat{\gamma}_{2j+1} +h\hat{\gamma}_{2j-1}\hat{\gamma}_{2j}],
\end{equation}
is obtained from the original one in Eq.~\eqref{eq:majorana_Ham} by shifting the Majorana lattice by one site. This translation exchanges the two Majorana couplings in Eq.~\eqref{eq:majorana_Ham} and leads to the same duality relation, $\hat{H}(h)=hH(1/h)$, found in the spin representation. The self-dual point $h=1$ is therefore the point at which the Hamiltonian is translationally invariant in the Majorana representation.

The quadratic fermionic Hamiltonian~\eqref{eq:majorana_Ham} can be diagonalized by means of a Bogoliubov transformation, which introduces a new set of fermionic (Dirac) quasiparticle operators $\eta_k$. In terms of these operators, the Hamiltonian takes the diagonal form
\begin{equation}
H(h)=\sum_k \epsilon_k(h) \eta_k^\dagger \eta_k,
\end{equation}
where $\epsilon_k(h)=\sqrt{1+h^2-2h\cos k}$ is the dispersion relation of the Bogolioubov modes. The ground state of the infinite chain is therefore the Bogoliubov vacuum $\ket{\Omega(h)}$, defined by 
$\eta_k\ket{\Omega(h)}=0$ for all $k$.

To compute the KW entanglement asymmetry of the ground state $\ket{\Omega(h)}$ we need the corresponding KW dual $\ket{\hat{\Omega}(h)}$, defined by Eq.~\eqref{eq:kw_tr}. Since $\ket{\Omega(h)}$ is the ground state of a quadratic fermionic Hamiltonian, it is a Gaussian state and satisfies Wick's theorem. 
Fermionic Gaussian states are completely characterized by their two-point correlation functions~\cite{peschel-03}.
Consequently, the dual state is specified by the relations
\begin{equation}\label{eq:KW_dual_gs}
\bra{\hat{\Omega}(h)}\hat{\gamma}_j\hat{\gamma}_{j'}\ket{\hat{\Omega}(h)}=\bra{\Omega(h)}\gamma_j\gamma_{j'}\ket{\Omega(h)}
\end{equation}
and is therefore also Gaussian.
As a result, the KW entanglement asymmetry can be computed entirely from the corresponding correlation matrices, defined as
\begin{equation}
\Gamma_{jj'}=\bra{\Omega(h)}
\boldsymbol{\gamma}_j^T
\boldsymbol{\gamma}_{j'} \ket{\Omega(h)}-\delta_{jj'}, \quad  \hat{\Gamma}_{jj'}=\bra{\hat{\Omega}(h)}
\boldsymbol{\gamma}_j^T
\boldsymbol{\gamma}_{j'}\ket{\hat{\Omega}(h)}-\delta_{jj'},
\end{equation}
where $\boldsymbol{\gamma}_j=(\gamma_{2j-1}, \gamma_{2j})$. In particular, the calculation of the R\'enyi entanglement asymmetry~\eqref{eq:renyi_EA} requires evaluating traces of products of Gaussian density matrices in our case. For a set of Gaussian states $\rho_1, \rho_2, \dots, \rho_n$ with correlation matrices $\Gamma_1, \Gamma_2, \dots, \Gamma_n$, these traces can be expressed as~\cite{fc-10} (see also \cite{bb-69})
\begin{equation}\label{eq:tr_det_gauss}
{\rm Tr}(\rho_1\rho_2\cdots\rho_n)=\sqrt{\prod_{j=1}^n\det\left(\frac{I-\Gamma_j}{2}\right)\det\left(I+\prod_{j=1}^n \frac{I+\Gamma_j}{I-\Gamma_j}\right)}.
\end{equation}
For example, for the case $n=2$, 
\begin{equation}
\Delta S_A^{(2)} = 2\log2 + \log{\rm Tr}\rho_A^2 - \log{\rm Tr} (\rho_A^2 + \hat{\rho}_A^2 + 2\rho_A\hat{\rho}_A),
\label{DS22}
\end{equation}
with $\rho_A$ and $\hat{\rho}_A$ the reduced density matrices of $\ket{\Omega(h)}$ and $\ket{\hat{\Omega}(h)}$ to subsystem $A$, respectively, Eq.~\eqref{eq:tr_det_gauss} gives 
\begin{equation}\label{eq:tr_det_renyi_2}
{\rm Tr}(\rho_A^2)=\sqrt{\det\left(\frac{I+\Gamma_A^2}{2}\right)}, \quad {\rm Tr}(\hat{\rho}_A^2)=\sqrt{\det\left(\frac{I+\hat{\Gamma}_A^2}{2}\right)}, \quad {\rm Tr}(\hat{\rho}_A\hat{\rho}_A)=\sqrt{\det\left(\frac{I+\Gamma_A\hat{\Gamma}_A}{2}\right)},
\end{equation}
where the subscript $A$ indicates that the correlation matrices are restricted to $A$. These expressions form the basis of our analysis: they allow for an efficient numerical evaluation of the KW entanglement asymmetry and provide the starting point for the analytical results presented in the following sections.
Notice, however, that although both $\rho_A$ and $\hat{\rho}_A$ are Gaussian, their sum is generally not. An alternative route, which we shall not explore here, is to define the symmetrized state directly within the manifold of Gaussian states. Such a Gaussian symmetrization would preserve the Gaussian structure and thus considerably simplify the evaluation of the entanglement asymmetry~\cite{tc-26}.

The correlation matrix $\Gamma$ of the ground state has the explicit form \cite{vlrk-03}
\begin{equation}\label{eq:Gamma_gs}
\Gamma_{jj'}=\int_{-\pi}^\pi\frac{dk}{2\pi}e^{ik(j-j')}\mathcal{G}(k),\quad  \mathcal{G}(k)=
\begin{pmatrix} 0 & -ie^{i\Delta_k(h)}\\ 
ie^{-i\Delta_k(h)} & 0\end{pmatrix},
\end{equation}
where $\Delta_k(h)$ is the Bogoliubov angle that diagonalizes the Hamiltonian~\eqref{eq:majorana_Ham}
\begin{equation}
\cos\Delta_k(h)=\frac{h-\cos k}{\epsilon_k(h)}, \quad \sin\Delta_k(h)=\frac{\sin k}{\epsilon_k(h)}.
\label{bog}
\end{equation}
Combining Eqs.~\eqref{eq:KW_dual_gs} and~\eqref{eq:KW_majorana}, we find that the correlation matrix $\hat{\Gamma}$ of the dual state has the same structure as $\Gamma$, with the replacement $\Delta_k(h)\mapsto \hat{\Delta}_k(h)=\pi-k-\Delta_k(h)$ in Eq.~\eqref{eq:Gamma_gs}.  Using the identity $\epsilon_k(h)=h\epsilon_k(1/h)$, this relation can be rewritten as $\hat{\Delta}_k(h)=\Delta_k(1/h)$. Therefore, the KW dual state $\ket{\hat{\Omega}(h)}$ coincides with the Bogoliubov vacuum of the Ising Hamiltonian at the dual coupling $1/h$, namely the ground state of $H(1/h)$.

We stress that, for $h<1$, the Gaussian ground state associated with the correlation matrix~\eqref{eq:Gamma_gs} is not one of the two ferromagnetic symmetry-broken ground states. Instead, it is their $\mathbb{Z}_2$-symmetric linear combination, which continuously approaches the equal-weight superposition in Eq.~\eqref{DS+-} as $h\to0$. This subtle but important point will be crucial for the analysis presented below.

The second scenario we consider is a quantum quench to the critical self-dual point. Specifically, the chain is initially prepared in the ground state of the quantum Ising Hamiltonian with transverse field $h=h_0\neq 1$. At time $t=0$, the magnetic field is suddenly quenched to the critical value $h=1$, so that the system evolves unitarily according to
\begin{equation}\label{eq:quench}
\ket{\Psi(t)}=e^{-itH(h=1)}\ket{\Omega(h_0)}.
\end{equation}
Although the full system evolves unitarily and does not relax to a stationary state, the reduced density matrix of any finite subsystem approaches, in the long-time limit, a stationary state described by a generalized Gibbs ensemble (GGE)~\cite{fe-16,vr-16}. Using the KW entanglement asymmetry, we will monitor the evolution of the KW duality after the quench and determine whether the stationary state of subsystem $A$ is self-dual, signaling the emergence of a local KW symmetry at long times. 
As in the equilibrium case, computing the asymmetry requires the KW dual $\ket{\hat{\Psi}(t)}$ of the time-evolved state~\eqref{eq:quench}. Given that the post-quench Hamiltonian $H(h=1)$ is self-dual, the KW dual of the evolved state is obtained by evolving the KW dual of the initial state with the same Hamiltonian,
\begin{equation}\label{eq:time_ev_dual}
\ket{\hat{\Psi}(t)}=e^{-itH(h=1)}\ket{\hat{\Omega}(h_0)}.
\end{equation}

Since the time evolution is generated by a quadratic Hamiltonian, the state $\ket{\Psi(t)}$ remains Gaussian at all times and is completely determined by its two-point correlation functions. 
The same holds for the KW dual state $\ket{\hat{\Psi}(t)}$. Therefore, the calculation of the KW entanglement asymmetry proceeds exactly as in the equilibrium case, using the corresponding time-dependent correlation matrices. For the state $\ket{\Psi(t)}$, it has been derived in Refs.~\cite{cc-05, fc-08} and reads
\begin{equation}\label{eq:time_ev_Gamma}
\Gamma_{jj'}(t)=\int_{-\pi}^{\pi}\frac{dk}{2\pi}e^{-ik(j-j')}\mathcal{G}(k, t),\quad 
		\mathcal{G}(k, t) = \begin{pmatrix} -f_{k}(t) & -g_k(t) \\
	 -g_{-k}(t) & f_k(t)
	\end{pmatrix},
\end{equation}
with
\begin{eqnarray}
	g_k(t) &=& -i e^{i\Delta_k(h_0)}(\cos\Theta_k + i \sin\Theta_k \cos(2\varepsilon_k(h) t)),\\
	f_k(t) &=& -\sin\Theta_{k} \sin(2\varepsilon_k(h)t).
\end{eqnarray}
The angle $\Theta_k$ is the difference between the Bogoliubov angles $\Delta_k(h_0)$ and $\Delta_k(h=1)$ that diagonalize the pre- and post-quench Hamiltonians,  
\begin{eqnarray}
    \cos\Theta_k &=& \frac{hh_0 - (h+h_0)\cos k + 1}{\varepsilon_k(h)\varepsilon_{k}(h_0)},\\
	\sin\Theta_k &=& -\frac{(h_0 - h) \sin k}{\varepsilon_k(h)\varepsilon_{k}(h_0)},
\end{eqnarray}
where, in our case, $h=1$. Given Eq.~\eqref{eq:time_ev_dual}, the correlation matrix $\hat{\Gamma}(t)$ of the KW dual of the time-evolved state has the same form as $\Gamma(t)$ with the initial Bogoliubov angle $\Delta_k(h_0)$ replaced by the transformed one $\hat{\Delta}_k(h_0)=\Delta_k(1/h_0)$. Hence, the KW transformation amounts to replacing $h_0$ by its dual value $1/h_0$ in Eq.~\eqref{eq:time_ev_Gamma}. 

With the machinery introduced above, we study the KW entanglement asymmetry of the Ising chain ground state in Sec.~\ref{sec4} and analyze its time evolution after a quantum quench to the self-dual point in Sec.~\ref{sec5}.

\section{Kramers-Wannier asymmetry in the ground state of the quantum Ising chain}
\label{sec4}

In this section, we investigate the behavior of the KW entanglement asymmetry in the ground state of the transverse-field Ising chain away from criticality. Throughout, we focus on the second Rényi KW entanglement asymmetry, $\Delta S_A^{(2)}$, introduced in the previous section in Eq. \eqref{DS22}, as it admits an efficient analytical and numerical treatment. For completeness, we reproduce its definition below:
\begin{equation}
\Delta S_A^{(2)} = 2\log2 + \log{\rm Tr}\rho_A^2 - \log{\rm Tr} (\rho_A^2 + \hat{\rho}_A^2 + 2\rho_A\hat{\rho}_A).
\label{DS2}
\end{equation}
At the self-dual point, we have $\rho_A=\hat{\rho}_A$ and clearly $\Delta S_A^{(2)}=0$, as expected.

 \begin{figure}[t]
\includegraphics[width = .5\textwidth]{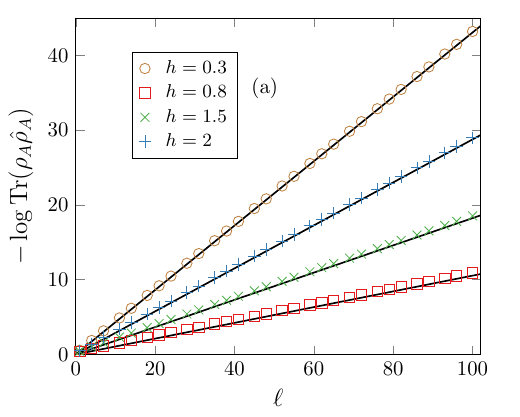}	
\includegraphics[width = .5\textwidth]{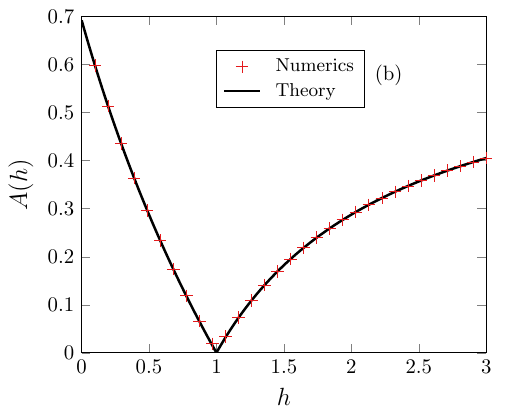}	
\caption{(a) ${\rm Tr} (\rho_A\hat{\rho}_A)$ as function of the subsystem size $\ell$ for the ground state of the quantum Ising chain for different magnetic fields $h$, showing the exponential decay.  The symbols are the exact values, computed using the corresponding formula in Eq.~\eqref{eq:tr_det_renyi_2}. The solid lines represent the asymptotic behavior in Eqs.~\eqref{tr_rho_rhohat_general}-\eqref{exponential_decrease_trpphat}, neglecting the subleading terms.  (b) The symbols are obtained from linear fits to the exact numerical data while the solid curve corresponds to the prediction of Eq.~\eqref{exponential_decrease_trpphat}.}
\label{exp_decrease}
\end{figure}

Away from criticality, we evaluate separately the three traces entering Eq.~\eqref{DS2}. We first consider $\mathrm{Tr}\,\rho_A\hat{\rho}_A$, which decays exponentially with the subsystem size, as shown in Fig.~\ref{exp_decrease}(a).
This behavior can be understood analytically using the Widom-Szegő theorem~\cite{WIDOM1974284, amvc-23}. Since both $\Gamma$ and $\hat{\Gamma}$ are block-Toeplitz matrices, the theorem provides the asymptotic behavior of the determinants appearing in Eq.~\eqref{DS2} in the limit of large subsystem size. The full derivation is reported in Appendix~\ref{app:trrhohat}; below we quote only the final expression
\begin{equation}
\log\Tr(\rho_A\hat{\rho}_A) = -A(h) \ell + O(1),
\label{tr_rho_rhohat_general}
\end{equation}
with
\begin{equation}
 	A(h) = \int_{-\pi}^{\pi} \frac{dk}{4\pi} (\log2 - \log[1 + \cos(\Delta_k - \hat\Delta_k)]) 	\equiv  \int_{-\pi}^{\pi} \frac{dk}{2\pi} f(k), \label{exponential_decrease_trpphat}
 \end{equation} 
where $\Delta_k$ and $\hat\Delta_k$ are the  Bogoliubov angles for $h$ and $1/h$ in Eq.~\eqref{bog}; we defined $f(k)$ for future use. 
The subleading terms are not relevant for what follows. 
A comparison between the numerics and Eq. \eqref{exponential_decrease_trpphat} is provided by Fig. \ref{exp_decrease}(b), and the agreement is excellent. Notice that $A\to 0$ as $h\to1$, signaling,  once again, the finiteness of the overlap at criticality. 

Conversely, it is well known that $\Tr\rho_A^2$ and $\Tr \hat{\rho}_A^2$  rapidly saturate to a finite value as $\ell$ increases. For both quantities, their exact asymptotic value can be inferred from the corner transfer matrix results presented in Refs. \cite{cc-04,Peschel-04,ccp-10,mac-17} (see also \cite{fik-08} for a different approach).
We quote the final result: 
\begin{equation}
  - \log	\Tr \rho_A^2 = \begin{cases}\parbox[t]{.7\textwidth}
  {$\displaystyle2\sum_{j \ge 1} [2\log(1 + e^{-2j \varepsilon}) - \log(1 + e^{-4j \varepsilon})]  + \log2$, \qquad\, \text{if $h < 1$},} \\ \\
  \parbox[t]{.5\textwidth} {$\displaystyle2\sum_{j \ge 0} [2\log(1 + e^{-(2j+1)\varepsilon}) - \log(1 + e^{-2(2j+1)\varepsilon})]$}, \qquad \;\; \text{if $h > 1$},
  \end{cases} 
  \label{CTM_purity}
\end{equation} 
with $\varepsilon$ being \cite{cc-04} 
\begin{equation}
\varepsilon = \pi \frac{I(\sqrt{1-k^2})}{I(k)}, \quad k = \begin{cases}|h|, &\text{if $h < 1$},\\ 1/|h|, &\text{if $h > 1$,}\end{cases}
\end{equation}
and $I(x)$ is the complete elliptic integral of the first kind.  As $h \to 1$, this theoretical expression diverges as expected. The result for $\Tr \hat\rho_A^2$ is simply obtained by replacing $h$ with $1/h$. The appearance of the additive $\log2$ term for $h<1$ is a manifestation of the Schrödinger-cat nature of the Gaussian ground state. Indeed, as anticipated, this state is the $\mathbb{Z}_2$-symmetric superposition of the two ferromagnetic symmetry-broken ground states. 

\begin{figure}[t]
\includegraphics[width = .5\textwidth]{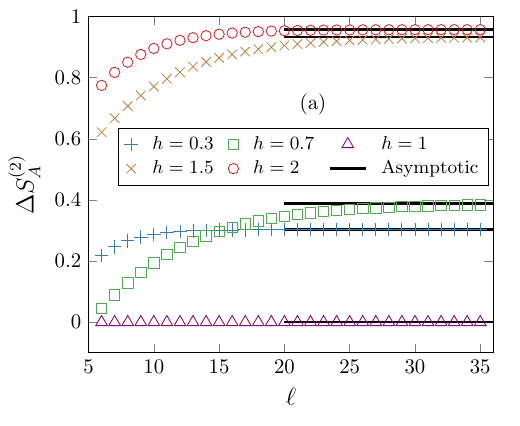}	
\includegraphics[width = .5\textwidth]{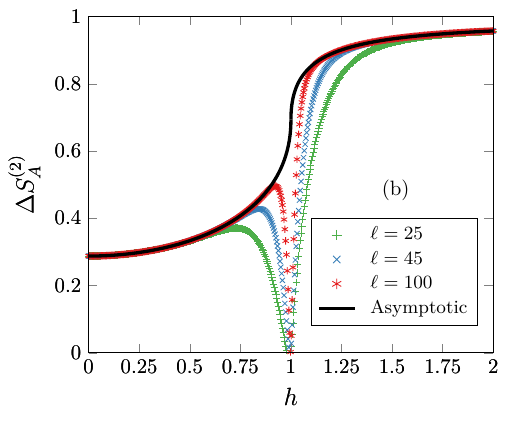}
\caption{(a) R\'enyi-2 KW entanglement asymmetry $\Delta S_A^{(2)}$ as a function of the subsystem size $\ell$ in the ground state of the quantum Ising chain for different magnetic fields $h$. (b) $\Delta S_A^{(2)}$ as a function of $h$ for $\ell = 25,45,100$.  In both panels, the symbols show the exact numerical values obtained from the correlation matrices at finite $\ell$ using Eq.~\eqref{eq:tr_det_renyi_2}, while the solid curves correspond to the asymptotic expression~\eqref{entanglement_asymmetry_simple}, with the traces evaluated from Eq.~\eqref{CTM_purity}  in the limit $\ell\to\infty$. }
\label{DeltaS_behavior}
\end{figure}

We now have all the ingredients needed to understand the behavior of the second Rényi KW entanglement asymmetry away from the self-dual point. The numerical results are reported in Fig.~\ref{DeltaS_behavior}(a). They show that $\Delta S_A^{(2)}$ approaches a finite asymptotic value as the subsystem size $\ell$ is increased. As expected, the convergence becomes slower upon approaching criticality, reflecting the divergence of the correlation length, so that progressively larger subsystem sizes are required to observe saturation as $h\to1$.
The approach to the asymptotic value is entirely governed by the overlap term $\Tr\rho_A\hat{\rho}_A$, whose exponential decay with $\ell$ was established in Eq.~\eqref{tr_rho_rhohat_general}. Consequently, in the limit of large subsystems this contribution becomes negligible, and the saturation value of the entanglement asymmetry for $h\neq1$ is obtained simply by dropping the term $\Tr \rho_A\hat{\rho}_A$ from Eq.~\eqref{DS2}. We thus arrive at 
\begin{equation}
	\Delta S_A^{(2)} \stackrel{\ell\to\infty} {\longrightarrow}2\log2 + \log\Tr\rho_A^2 - \log\Tr(\rho_A^2 + \hat{\rho}_A^2),
\label{entanglement_asymmetry_simple}
\end{equation}
with $\Tr\rho_A^2, \Tr\hat{\rho}_A^2$ being given by Eq.~\eqref{CTM_purity}. 
This is our final analytic expression for the KW entanglement asymmetry and it shows that, away from criticality, its asymptotic value is completely determined by the purities of the reduced density matrix and of its KW dual. The exponentially small overlap between the two reduced density matrices only gives rise to finite-size corrections, which vanish in the thermodynamic limit of large subsystems.

The comparison between the analytical prediction~\eqref{entanglement_asymmetry_simple} and the exact numerical results is presented in Fig.~\ref{DeltaS_behavior}(b). As expected, the numerical data show that the second Rényi KW entanglement asymmetry vanishes exactly at the self-dual point, $\Delta S_A^{(2)}(h=1)=0$, reflecting the exact Kramers-Wannier invariance of the critical ground state. In contrast, the asymptotic expression~\eqref{entanglement_asymmetry_simple} does not vanish at $h=1$. This apparent discrepancy originates from the approximation leading to Eq.~\eqref{entanglement_asymmetry_simple}, where the overlap term $\Tr\rho_A\hat{\rho}_A$ was neglected. While this approximation is justified throughout the gapped phases because the overlap decays exponentially with the subsystem size, it breaks down precisely at the critical point. Indeed, at $h=1$, the overlap is no longer exponentially suppressed and becomes of the same order as (indeed, equal to) the two purity terms entering Eq.~\eqref{DS2}. Their cancellation is therefore essential to recover the exact result $\Delta S_A^{(2)}=0$.
This observation immediately clarifies how the asymptotic behavior is approached. The validity of Eq.~\eqref{entanglement_asymmetry_simple} depends on both the subsystem size and the distance from criticality. At any finite $\ell$, the entanglement asymmetry therefore exhibits a pronounced dip around the critical point, where the neglected overlap becomes significant and drives the asymmetry to zero. As the subsystem size increases, this critical region shrinks, and the asymptotic expression becomes accurate over an increasingly broad range of magnetic fields. The width of the crossover region can be estimated by requiring that the overlap is no longer negligible, namely by imposing
$A\ell\sim1$, where $A$ is the decay rate appearing in Eq.~\eqref{tr_rho_rhohat_general}. Since $A$ vanishes at the self-dual point, this condition determines the width of the critical region where finite-size effects remain relevant. In the limit $\ell\to\infty$, the width of this region vanishes, so that the asymptotic result~\eqref{entanglement_asymmetry_simple} is recovered everywhere except exactly at the critical point, where the asymmetry remains identically zero by symmetry.
Interestingly, a similar dip has also been reported in the entanglement asymmetry associated with spatial inversion symmetry~\cite{hey-26}.

It is also instructive to examine the value of $\Delta S_A^{(2)}$ at the two limiting points of the phase diagram $h=0$ and $h=\infty$. From Eq.~\eqref{CTM_purity}, the corresponding subsystem purities are
\begin{equation}
\Tr\rho_A^2(h=0)=\frac{1}{2},\qquad
\Tr\rho_A^2(h=\infty)=1.
\end{equation}
Substituting these values into Eq.~\eqref{entanglement_asymmetry_simple}, we obtain
\begin{equation}
\Delta S_A^{(2)}(h=0)=\log\frac{4}{3},\qquad
\Delta S_A^{(2)}(h=\infty)=\log\frac{8}{3}.
\end{equation}
These values coincide with the second Rényi KW entanglement asymmetries of the conformal boundary states derived in Ref.~\cite{benini-25} and reviewed in Sec.~\ref{sec2} (cf. Eqs.~\eqref{DS+-} and~\eqref{DSf}). 
It is remarkable that the calculation performed on the microscopic lattice model reproduces boundary conformal field theory, although the identification is more subtle (see Sec.~\ref{sec2}).

Finally, although it has no direct physical significance, it is interesting to analyze the behavior of the asymptotic expression~\eqref{entanglement_asymmetry_simple} in the limit $h\to1$. A straightforward asymptotic analysis of Eq.~\eqref{entanglement_asymmetry_simple} shows that the asymptotic curve approaches the finite value $\log2$.
At the same time, its derivative diverges as the critical point is approached. Thus, while the asymptotic expression remains continuous at $h=1$, it develops a non-analyticity. Both features are clearly visible in Fig.~\ref{DeltaS_behavior}(b).
Of course, this limiting behavior should not be interpreted as the physical value of the entanglement asymmetry at criticality, since Eq.~\eqref{entanglement_asymmetry_simple} is derived by neglecting the overlap term $\Tr\rho_A\hat{\rho}_A$, an approximation that breaks down precisely at the self-dual point.

It is worth noticing that the KW entanglement asymmetry displays some qualitative similarities with the standard $\mathbb{Z}_2$ entanglement asymmetry. In both cases, the asymmetry saturates to an $O(1)$ value in the symmetry-broken phase~\cite{fac-23b,cm-23}. The analogy, however, stops there: whereas the $\mathbb{Z}_2$ asymmetry always saturates to the universal value $\log2$, irrespective of the transverse field $h<1$, the Kramers–Wannier asymmetry retains a nontrivial dependence on $h$.

\subsection{Higher R\'enyi Kramers-Wannier asymmetry and analytic continuation}

In this subsection, we extend the equilibrium analysis to the Rényi KW entanglement asymmetry with arbitrary integer Rényi index $n$. We start from the definition
\begin{equation}
\Delta S_A^{(n)}
=
S^{(n)}\!\left(\frac{\rho_A+\hat\rho_A}{2}\right)
-
S^{(n)}(\rho_A).
\end{equation}

Away from the self-dual point, the reduced density matrices $\rho_A$ and $\hat\rho_A$ become asymptotically orthogonal as the subsystem size increases, as witnessed by the exponential decay of the overlap $\Tr\rho_A\hat\rho_A$. As a consequence, all mixed products appearing in the expansion of $(\rho_A+\hat\rho_A)^n$ are exponentially suppressed in $\ell$. This can be shown explicitly starting from the Gaussian expression~\eqref{eq:tr_det_gauss} and applying the generalized Widom-Szegő theorem reported in Ref.~\cite{amvc-23}. To leading order in the subsystem size, one therefore has
\begin{equation}
\Tr\left(\frac{\rho_A+\hat\rho_A}{2}\right)^n
\simeq
\frac{\Tr\rho_A^n+\Tr\hat\rho_A^n}{2^n}.
\end{equation}
It follows that, for $h\neq1$ and sufficiently large $\ell$,
\begin{equation}
\Delta S_A^{(n)}
\simeq
\frac{1}{1-n}
\log\left[
\frac{\Tr\rho_A^n+\Tr\hat\rho_A^n}
{2^n\Tr\rho_A^n}
\right].
\label{DeltaSn_general}
\end{equation}

The Rényi traces $\Tr\rho_A^n$ and $\Tr\hat\rho_A^n$ are known exactly from the corner transfer matrix approach~\cite{cc-04,Peschel-04,ccp-10}. They are given by
\begin{equation}
-\log\Tr\rho_A^n=
\begin{cases}
\displaystyle
2\sum_{j\ge1}
\left[
n\log(1+e^{-2j\varepsilon})
-
\log(1+e^{-2nj\varepsilon})
\right]
+(n-1)\log2,
& h<1,
\\[0.6cm]
\displaystyle
2\sum_{j\ge0}
\left[
n\log(1+e^{-(2j+1)\varepsilon})
-
\log(1+e^{-n(2j+1)\varepsilon})
\right],
& h>1,
\end{cases}
\label{CTM_n}
\end{equation}
where the corresponding expression for $\Tr\hat\rho_A^n$ is obtained by the duality transformation $h\leftrightarrow1/h$.

The derivation above shows that the only genuinely new ingredient required to compute the Rényi KW entanglement asymmetry is the exponential suppression of the mixed products between $\rho_A$ and $\hat\rho_A$. Once this property is established, the computation for arbitrary integer $n$ follows directly from the known corner transfer matrix expressions for the Rényi entropies.

\begin{figure}[t]
\includegraphics[width = .5\textwidth]{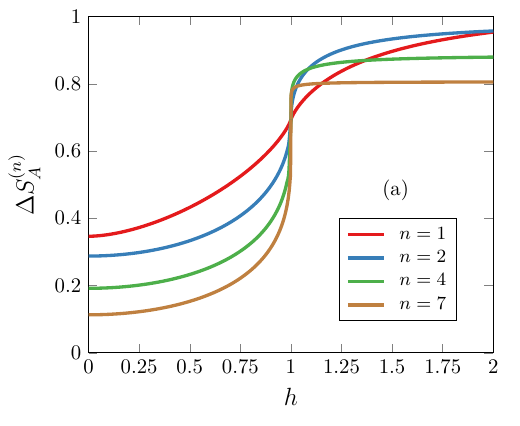}	
\includegraphics[width = .5\textwidth]{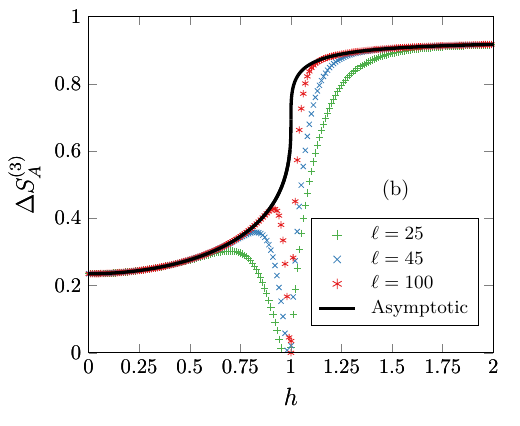}
\caption{R\'enyi KW entanglement asymmetry for several R\'enyi indices $n$ as function of the magnetic field $h$ in the ground state of the quantum Ising chain. (a) Asymptotic behavior in Eqs.~\eqref{DeltaSn_general}-\eqref{CTM_n}, including the analytic continuation for $n=1$. 
(b) Comparison with exact numerics at finite $\ell$ for $n=3$, showing the typical dip at criticality, signaling the restoration of symmetry not captured by Eq.~\eqref{DeltaSn_general} as in Fig. \ref{DeltaS_behavior} for $n=2$.}
\label{DeltaS_renyi}
\end{figure}

In Fig.~\ref{DeltaS_renyi}(a) we report the asymptotic prediction~\eqref{DeltaSn_general} as a function of the transverse field for several Rényi indices. This expression describes the massive-regime behavior away from the self-dual point, where the mixed products between $\rho_A$ and $\hat\rho_A$ are exponentially suppressed. As in the second Rényi case, it should not be expected to reproduce the immediate vicinity of $h=1$, where the overlap terms become non-negligible and are essential to recover the exact vanishing of the asymmetry at the self-dual point. However, in this critical region, the asymptotic curves becomes sharper as increasing $n$ and they all cross at $\log2$ at the critical point $h=1$.
The limiting values for $h=0$ and $h=\infty$ match always the ones in Eqs.~\eqref{DS+-} and~\eqref{DSf}, respectively.
For $n\to\infty$, the KW asymmetry becomes a step function jumping from $0$ to $\log2$ at the critical point.

As an explicit benchmark, in Fig.~\ref{DeltaS_renyi}(b) we compare the asymptotic prediction for the third Rényi KW entanglement asymmetry with exact numerical calculations. Away from the critical region, the agreement is excellent. Close to $h=1$, however, the numerical data display a pronounced dip that is absent from Eq.~\eqref{DeltaSn_general}. This discrepancy has the same origin as for $n=2$: the asymptotic CTM expression neglects mixed products involving both $\rho_A$ and $\hat\rho_A$, which become relevant near the self-dual point. These terms restore the exact result $\Delta S_A^{(n)}=0$ at $h=1$ for any finite subsystem size.
Upon increasing $\ell$, the region in which the mixed terms are relevant shrinks, and the numerical data approach the massive-regime prediction~\eqref{DeltaSn_general} everywhere except in an increasingly narrow neighborhood of the self-dual point, exactly as for $n=2$. Thus the finite-$\ell$ dip should be understood as a crossover effect controlled by the competition between the subsystem size and the vanishing decay rate of the overlap close to criticality.

Finally Eqs. \eqref{DeltaSn_general} and \eqref{CTM_n} can be easily analytically continued at $n\to1$ to obtain the von Neumann KW asymmetry, the final result being 
\begin{equation}
\Delta S_A(h)
=
\log 2
-
\frac{1}{2}
\left[
S_A(h)-S_A(1/h)
\right],
\qquad h\neq1,
\end{equation}
with
\begin{equation}
S_A(h)=-\Tr\rho_A\log\rho_A=
\begin{cases}
\displaystyle
\log2
+
2\sum_{j\ge1}
\left[
\log(1+e^{-2j\varepsilon})
+
\frac{2j\varepsilon}{1+e^{2j\varepsilon}}
\right],
& h<1,
\\[0.7cm]
\displaystyle
2\sum_{j\ge0}
\left[
\log(1+e^{-(2j+1)\varepsilon})
+
\frac{(2j+1)\varepsilon}{1+e^{(2j+1)\varepsilon}}
\right],
& h>1,
\end{cases}
\label{CTM_vN}
\end{equation}
which is also shown in Fig.~\ref{DeltaS_renyi}(a).

\section{Kramers-Wannier asymmetry after a  quantum quench to the self-dual point}\label{sec5}

In this section, we investigate the nonequilibrium dynamics of the KW entanglement asymmetry following a quantum quench. Specifically, we consider systems initially prepared in the ground state of the transverse-field Ising chain at $h=h_0$, which are then suddenly quenched to the critical self-dual point $h=1$. Our main object of interest is the time evolution of the second Rényi Kramers-Wannier entanglement asymmetry, $\Delta S_A^{(2)}(t)$.
The first question we address is whether the Kramers-Wannier symmetry is dynamically restored during the unitary evolution, namely whether $\Delta S_A^{(2)}(t\to\infty)=0$. At first sight, one might expect the answer to be affirmative, since the post-quench Hamiltonian is exactly invariant under the Kramers-Wannier duality. However, this conclusion is far from obvious. Unlike in equilibrium, no general theorem guarantees the restoration of a symmetry during nonequilibrium dynamics. In fact, even for ordinary invertible symmetries, there are well-established examples in which a quench to a symmetric Hamiltonian fails to restore the symmetry asymptotically~\cite{amvc-23, cma-24, yca-25, dgtm-25}. Whether an analogous phenomenon can occur for a non-invertible symmetry such as the Kramers-Wannier duality is therefore an open and highly nontrivial question.

A second motivation for our analysis is the possibility of observing a quantum Mpemba effect~\cite{acm-25,c-26}. In the present context, this would correspond to a situation in which states with a larger initial KW entanglement asymmetry restore the non-invertible symmetry faster than states that are initially closer to the symmetric configuration. Such an apparently counterintuitive behavior has recently attracted considerable attention in the context of symmetry restoration for ordinary symmetries and other nonequilibrium phenomena. Here we investigate whether an analogous effect exists for the Kramers-Wannier duality, and, if so, identify the physical mechanism responsible for its emergence.

\begin{figure}
\includegraphics[width = .5\textwidth]{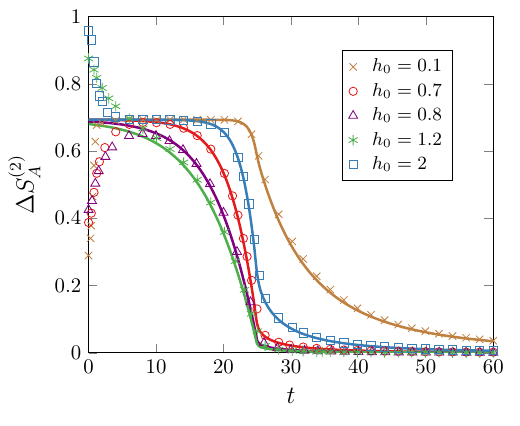}
\includegraphics[width = .5\textwidth]{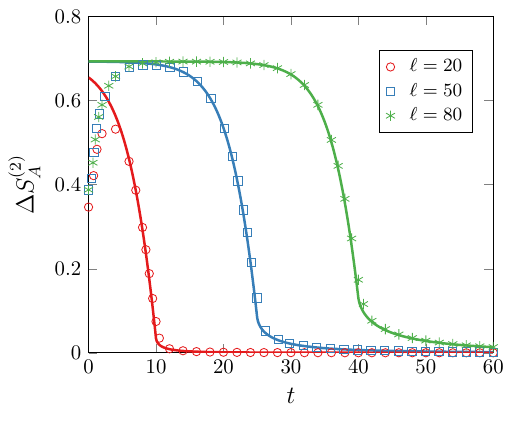}
\caption{(a) Time evolution of R\'enyi-2 KW entanglement asymmetry $\Delta S_A^{(2)}(t)$ for a quench to the critical point of the quantum Ising chain for different values of the initial magnetic field $h$.
The symbols are the exact numerical results for $\ell = 50$, computed from the correlation matrix using Eq.~\eqref{eq:tr_det_renyi_2}.
The full lines represent the asymptotic results from the quasiparticle picture.  (b) Time evolution of $\Delta S_A^{(2)}(t)$ for the representative quench from $h_0=0.7$ and several subsystem sizes $\ell$ (symbols). The quasiparticle predictions are shown as solid lines. The figure clearly illustrates that the initial transient is an $O(1)$ effect, and is therefore not captured by the quasiparticle picture. It also shows how the intermediate plateau progressively develops as the subsystem size is increased.}
\label{Fig:quench}
\end{figure}

The entanglement asymmetry can be computed numerically using the correlation-matrix techniques described in Sec.~\ref{sec3}. The results are reported in Fig.~\ref{Fig:quench}(a) (symbols) for several values of the initial field $h_0$ in both the ordered and disordered phases, keeping the subsystem size fixed at $\ell=50$.
Even before attempting an analytical interpretation, the numerical data already reveal several interesting features. First, the asymmetry always goes to zero for large times, showing that generally the Kramers-Wannier symmetry is restored regardless of the initial value of $h_0$. 
Then the relaxation dynamics is rather rich and, in general, far from monotonic. In particular, for quenches originating in the ordered phase ($h_0<1$), the entanglement asymmetry exhibits a non-monotonic time dependence.
After an initial transient lasting a time of order unity, i.e., not scaling with the subsystem size $\ell$, as highlighted in Fig.~\ref{Fig:quench}(b), the asymmetry reaches an intermediate plateau at $\Delta S_A^{(2)}=\log2$,
for a broad range of initial values of $h_0$. The plateau is especially well developed for quenches starting sufficiently far from the critical point. As the initial field approaches the critical value, however, the plateau becomes progressively shorter and eventually disappears. This should not be interpreted as the absence of the plateau. Rather, for the subsystem size considered here, the asymmetry relaxes towards its asymptotic value before the plateau has enough time to develop. We therefore expect that, upon increasing $\ell$, the restoration dynamics will be delayed, making the plateau visible over a progressively wider range of initial conditions and, ultimately, for all values of $h_0$. For the representative quench with $h_0=0.7$, Fig.~\ref{Fig:quench}(b) illustrates how the intermediate plateau develops as the subsystem size is increased. While it is essentially absent for $\ell=20$, it becomes clearly visible for $\ell=50$ and is well developed and long-lived for $\ell=80$.

Another striking feature emerging from Fig.~\ref{Fig:quench}(a) is the occurrence of the quantum Mpemba effect. For example, the quench from $h_0=2$ relaxes faster than that from $h_0=0.7$, despite the fact that the former starts from a state with a larger KW entanglement asymmetry, i.e., one that initially breaks the symmetry more strongly.
However, the relation between the initial amount of symmetry breaking and the relaxation rate is clearly not universal. As illustrated in the figure, there are quenches from states initially with lower KW asymmetry that do not relax more slowly than all others. Thus, the occurrence of the quantum Mpemba effect cannot be inferred solely from the initial value of the entanglement asymmetry.
Determining which pairs of initial conditions exhibit a quantum Mpemba effect therefore requires a detailed analytical understanding of the long-time relaxation dynamics. In the following, we develop such an understanding by exploiting the quasiparticle picture \cite{cc-05,fc-08,ac-17,c-20}, which allows us to derive the asymptotic behavior of the entanglement asymmetry and identify the precise conditions under which the Mpemba effect arises.

\subsection{The quasi-particle picture for Kramers-Wannier asymmetry}
\label{QP}

In the quasiparticle picture, the dynamics of entanglement is interpreted as the result of the ballistic propagation of entangled pairs of quasiparticles created by the quench. This picture becomes asymptotically exact in the scaling limit $t,\ell\to\infty$ with the ratio $t/\ell$ held fixed.
As a paradigmatic example, the leading contribution to the second Rényi entropy is given by~\cite{fc-08}
\begin{equation}
	-\log\Tr\rho_A^2 = \int_{0}^{2\pi} \frac{dk}{2\pi} \min(\ell, 2|v_k| t )h_2(\cos\Theta_k),
	\label{trace_rho2_dyna}
\end{equation}
where 
\begin{equation}
	h_2(x) = \log\left[\left(\frac{1 + x}{2}\right)^2 + \left(\frac{1 - x}{2}\right)^2\right],
\end{equation}
and $v_k=\varepsilon_k'$ is the quasiparticle group velocity.
Eq.~\eqref{trace_rho2_dyna} admits a simple and intuitive interpretation. Since the initial state is not an eigenstate of the post-quench Hamiltonian, it acts as a homogeneous source of quasiparticle pairs. Each pair consists of quasiparticles with opposite momenta $\pm k$, which propagate ballistically with velocities $\pm v_k$. The two quasiparticles belonging to the same pair are entangled, whereas different pairs are assumed to be uncorrelated.
Each quasiparticle pair contributes an amount $h_2(\cos\Theta_k)$ to the second Rényi entropy, provided that one quasiparticle lies inside the subsystem $A$ while its partner lies outside. Pairs for which both quasiparticles are either inside $A$ or outside $A$ do not contribute to the entanglement between $A$ and its complement. It is then straightforward to show that the number of contributing pairs is precisely proportional to $\min(\ell,2|v_k|t)$, yielding Eq.~\eqref{trace_rho2_dyna}. 
Notice that, at $t=0$, the quasiparticle formula predicts a vanishing second Rényi entropy. This should not be interpreted as implying that the entropy itself is exactly zero. Rather, the quasiparticle picture only captures the leading extensive contribution, proportional to the subsystem size $\ell$. Thus, a vanishing result simply indicates the absence of an extensive term, while leaving room for subleading contributions, such as terms of order unity or even logarithmic corrections, the latter occurring, for example, when the initial state is critical \cite{cc-04}.
This observation will be particularly relevant in the following when discussing the Kramers-Wannier entanglement asymmetry. Indeed, while the quasiparticle picture correctly describes its scaling behavior at intermediate and long times, it does not capture the subleading contributions that determine its initial value. As a consequence, the quasiparticle prediction necessarily fails in the short-time regime, even though it becomes asymptotically exact in the scaling limit $t,\ell\to\infty$ with $t/\ell$ fixed.

The quasiparticle result above, Eq.~\eqref{trace_rho2_dyna}, immediately provides the time evolution of both
$\Tr\rho_A^2$ and $\Tr\hat{\rho}_A^2$. Therefore, the only ingredient missing to determine the KW entanglement asymmetry through Eq.~\eqref{DS2} is the mixed overlap $\Tr\rho_A\hat{\rho}_A$. We now argue that this quantity also admits a simple quasiparticle interpretation.
The starting point is the observation that both its initial and long-time extensive values are known exactly. At the initial time, Eqs.~\eqref{tr_rho_rhohat_general} and \eqref{exponential_decrease_trpphat} imply
\begin{equation}
\log\Tr\rho_A\hat{\rho}_A(t=0)
=
-\ell\int_0^{2\pi}\frac{dk}{2\pi}\,f(k),
\label{initial_trrhohat}
\end{equation}
where the function $f(k)$ is defined in Eq.~\eqref{exponential_decrease_trpphat}.
The asymptotic value for $t\to\infty$ is also known. Indeed, assuming that the Kramers-Wannier symmetry is dynamically restored at long times, one has
\begin{equation}
\Tr\rho_A\hat{\rho}_A(t\to\infty)
=
\Tr\rho_A^2(t\to\infty).
\end{equation}
The latter is given by the quasiparticle formula \eqref{trace_rho2_dyna}, which in the stationary regime yields
\begin{equation}
\log\Tr\rho_A\hat{\rho}_A(t\to\infty)
=
-\ell\int_0^{2\pi}\frac{dk}{2\pi}\,
h_2(\cos\Theta_k).
\label{final_trrhohat}
\end{equation}

These two limiting expressions naturally suggest a quasiparticle description for the full time evolution. In complete analogy with Eq.~\eqref{trace_rho2_dyna}, we assume that each momentum mode relaxes independently and that the fraction of quasiparticle pairs contributing at time $t$ is again described by the factor $\min(1,2|v_k|t/\ell)$. This leads to the conjecture
\begin{equation}
\log\Tr\rho_A\hat{\rho}_A(t)
=
-\ell\int_0^{2\pi}\frac{dk}{2\pi}\,
\min\!\left(1,\frac{2|v_k|t}{\ell}\right)
\bigl[h_2(\cos\Theta_k)-f(k)\bigr]
-\ell\int_0^{2\pi}\frac{dk}{2\pi}\,f(k).
\label{tr_rhorhohat_dyna}
\end{equation}
By construction, Eq.~\eqref{tr_rhorhohat_dyna} reproduces exactly both the initial condition,
Eq.~\eqref{initial_trrhohat}, and the stationary limit,
Eq.~\eqref{final_trrhohat}. Moreover, it has the same quasiparticle structure as the Rényi entropy, with the only difference that the contribution of each momentum mode is shifted by the function $f(k)$, which encodes the initial mismatch between $\rho_A$ and its KW transformed counterpart.

\begin{figure}[t]
  \centering
\includegraphics[width = .5\textwidth]{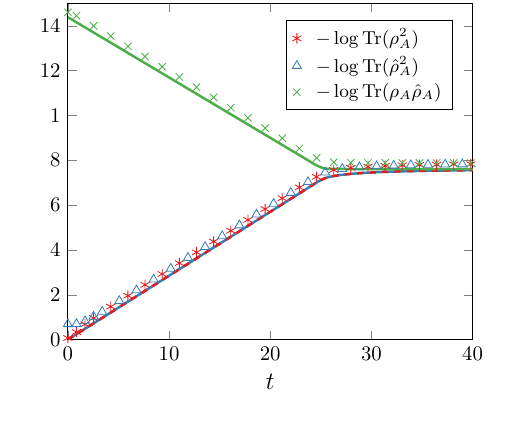}
\caption{Time evolution of the traces entering the R\'enyi-2 Kramers–Wannier entanglement asymmetry~\eqref{DS2} after a quench from $h_0 = 2$  to the critical point. Here the susbystem size is $\ell = 50$. The symbols are the exact numerical values, computed using the formulas in Eq.~\eqref{eq:tr_det_renyi_2}, while the curves represent the quasiparticle predictions in Eqs.~\eqref{trace_rho2_dyna} and~\eqref{tr_rhorhohat_dyna}. Observe that the quasiparticle prediction for $\Tr\rho_A^2$ and its KW dual $\Tr\hat\rho_A^2$ coincide for any initial magnetic field $h_0$. }
\label{quench_traces}
\end{figure} 

The prediction \eqref{tr_rhorhohat_dyna} is compared with the numerical data in Fig.~\ref{quench_traces}, where we find excellent agreement over the entire time evolution, providing strong evidence for the validity of the quasiparticle picture also for the mixed overlap $\Tr\rho_A\hat{\rho}_A$.

We now have all the ingredients needed to obtain an analytical prediction for the time evolution of the second Rényi KW entanglement asymmetry. Combining the three traces entering Eq.~\eqref{DS2}, we obtain the quasiparticle prediction for $\Delta S_A^{(2)}(t)$. The resulting curves are shown as full lines in Fig.~\ref{Fig:quench}(a), together with the exact numerical results.
The agreement is remarkably good. In particular, the quasiparticle prediction accurately captures the entire long-time dynamics, including the emergence of the intermediate plateau at $\Delta S_A^{(2)}=\log2$ and the subsequent relaxation towards zero, signaling the dynamical restoration of the Kramers-Wannier symmetry. Consequently, it also correctly reproduces the occurrence of the quantum Mpemba effect, including the ordering of the relaxation times for different initial values of $h_0$.
The agreement deteriorates only at short times, where the quasiparticle prediction does not reproduce the exact initial value of the entanglement asymmetry. As already anticipated above, this limitation is expected. Indeed, the quasiparticle picture employed here provides only the leading contribution in the scaling limit $t,\ell\to\infty$ with the ratio $t/\ell$ kept fixed. As a consequence, it neglects subleading corrections, which are responsible for the short-time transient and, in particular, for the precise value of $\Delta S_A^{(2)}(0)$. Nevertheless, these corrections rapidly become negligible, explaining the excellent agreement observed already at intermediate times.

The origin of the intermediate plateau admits a simple interpretation within the quasiparticle picture. First, one can verify, both analytically and numerically, that in the scaling regime
$\log\Tr\rho_A^2(t)=\log\Tr\hat{\rho}_A^2(t)$, 
as clear from the quasiparticle approach. Moreover, the overlap $\Tr\rho_A\hat{\rho}_A$ remains exponentially small at short times, since it is exponentially suppressed already at $t=0$ and only becomes appreciable at much later times. Consequently, in this regime one can neglect the overlap term in Eq.~\eqref{DS2}, obtaining
\begin{equation}
\Delta S_A^{(2)}(t)
\simeq
2\log2
+\log\Tr\rho_A^2(t)
-\log\!\left(2\Tr\rho_A^2(t)\right)
=
\log2.
\end{equation}
This immediately explains the appearance of the plateau at $\Delta S_A^{(2)}=\log2$.

The plateau persists up to times of order $t\simeq\ell/2$. This timescale has a simple quasiparticle interpretation. Indeed, the factor $\min(\ell,2|v_k|t)$ appearing in Eq.~\eqref{tr_rhorhohat_dyna} counts the number of entangled quasiparticle pairs shared between the subsystem $A$ and its complement. Since the quasiparticle velocities satisfy $|v_k|\le1$, this quantity grows linearly with time until the fastest quasiparticles have traversed the subsystem, namely for $t\lesssim\ell/2$. During this stage, the overlap $\Tr\rho_A\hat{\rho}_A$ remains negligible and the entanglement asymmetry stays pinned to the plateau value.
For $t\gtrsim\ell/2$, the dynamics enters a different regime. The contribution of the fastest quasiparticles has already saturated, and the various momentum modes progressively relax towards their stationary values (see also Fig.~\ref{quench_traces}). Correspondingly, the overlap $\Tr\rho_A\hat{\rho}_A$ is no longer negligible and starts increasing towards $\Tr\rho_A^2$. Since all three traces entering Eq.~\eqref{DS2} become equal in the stationary state, the KW entanglement asymmetry eventually decays to zero, signaling the dynamical restoration of the Kramers-Wannier symmetry.

\subsection{Conditions for the quantum Mpemba effect}

As already mentioned, a quantum Mpemba effect occurs when a state that initially breaks the symmetry more strongly restores it faster than another state that is initially closer to the symmetric configuration. For the KW entanglement asymmetry, this means that, given two initial states $a$ and $b$ satisfying
\begin{equation}
(\Delta S_A^{(2)})_a(0)>
(\Delta S_A^{(2)})_b(0),
\end{equation}
a quantum Mpemba effect takes place if there exists a time $t_M$ such that
\begin{equation}
(\Delta S_A^{(2)})_a(t)
<
(\Delta S_A^{(2)})_b(t),
\qquad
\forall\, t>t_M.
\end{equation}

For two quenches characterized by different initial values of the transverse field $h_0$, the first condition can be readily established from the equilibrium results of the previous section. Indeed, in the large-$\ell$ limit, the initial asymmetry is given by Eq.~\eqref{entanglement_asymmetry_simple}, so that the ordering of the initial values is known exactly.

To determine whether the second condition is fulfilled, one has to characterize the long-time relaxation. This can be achieved following exactly the same strategy developed for ordinary invertible symmetries in Refs.~\cite{rka-23,makc-23}. Within the quasiparticle picture, the asymptotic dynamics is governed by the slowest quasiparticles. Therefore, the long-time behavior can be extracted by expanding the quasiparticle integrals around the stationary points of the dispersion relation, where the quasiparticle velocity vanishes. The calculation is straightforward, although somewhat lengthy, and is presented in Appendix~\ref{app:LT}. The final result reads
\begin{equation}
\Delta S_A^{(2)}(t)
\underset{t\to\infty}{\sim}
\frac{\ell^4}{192\pi t^3}
\left(\frac{h_0-1}{h_0+1}\right)^2.
\end{equation}
The decay exponent is universal, while the dependence on the initial state is entirely encoded in the prefactor. As a function of $h_0$, this prefactor is monotonically decreasing throughout the ordered phase ($h_0<1$) and monotonically increasing throughout the disordered phase ($h_0>1$), attaining its minimum at the critical point.

Combining this result with the behavior of the initial asymmetry, shown in Fig.~\ref{DeltaS_behavior}(b), immediately yields the conditions for the occurrence of the quantum Mpemba effect. Away from the narrow critical region where finite-size effects become relevant, the initial asymmetry is a monotonically increasing function of $h_0$. Consequently, if both initial states belong to the ordered phase ($h_0<1$), the state that initially breaks the Kramers-Wannier symmetry more strongly always has a smaller asymptotic prefactor and therefore relaxes faster. In contrast, if both initial states lie in the disordered phase ($h_0>1$), the ordering of the prefactor is the same as that of the initial asymmetry, and no Mpemba effect can occur. Finally, for pairs of initial states belonging to opposite phases, the occurrence of the Mpemba effect depends on the specific values of the two initial fields and must be analyzed on a case-by-case basis, just by looking at the formulas presented above.

\section{Conclusions}
\label{sec6}

In this work we studied the KW entanglement asymmetry, a quantum-information measure that quantifies the breaking of the Kramers-Wannier duality at the level of reduced density matrices. This construction extends the notion of entanglement asymmetry from ordinary invertible symmetries to the paradigmatic example of a non-invertible symmetry, providing a novel tool to investigate dualities in quantum many-body systems.

We first analyzed the equilibrium properties of the transverse-field Ising chain. Away from criticality, we obtained an exact characterization of the  Rényi KW entanglement asymmetries using correlation-matrix techniques. We showed that it saturates to a finite value for large subsystem size and exhibits a pronounced dip at the self-dual point, whose width shrinks as the subsystem size increases. 

We then investigated the nonequilibrium dynamics following quenches to the critical point. We found that the KW entanglement asymmetry vanishes asymptotically, providing direct evidence for the dynamical restoration of the non-invertible symmetry. Furthermore, we generalized the quasiparticle picture to the mixed overlap $\Tr\rho_A\hat{\rho}_A$, obtaining analytical predictions for the entire long-time dynamics. These successfully reproduce the intermediate plateau, the asymptotic decay, and the emergence of a quantum Mpemba effect. In particular, we derived the precise conditions under which the Mpemba effect occurs, relating it to the asymptotic relaxation governed by the slowest quasiparticles.

Our results naturally suggest several directions for future investigation. On the equilibrium side, it would be interesting to extend the notion of entanglement asymmetry to other lattice realizations of non-invertible symmetries and to compare the resulting universal quantities with the predictions recently obtained within boundary conformal field theory~\cite{benini-25,bgvv-26}. Such a program could provide a systematic correspondence between lattice models and the boundary-CFT description of non-invertible symmetries.
An obvious extension is the study of lattice models realizing more general non-invertible symmetries, such as the three-state Potts chain and parafermionic quantum chains~\cite{afm-20}, where richer fusion categories are expected to produce qualitatively new behaviors of the entanglement asymmetry.

Another natural question concerns the crossover between the critical point, where the KW entanglement asymmetry vanishes identically, and the finite asymptotic values characterizing the massive phases. For the entanglement entropy, analogous crossover functions have been successfully obtained using the form-factor bootstrap approach in integrable quantum field theory~\cite{ccd-08,cd-09}. Extending these techniques to the KW entanglement asymmetry would require the characterization of the mixed overlap $\Tr\rho_A\hat{\rho}_A$ in terms of form factors, a challenging but potentially rewarding problem.

Another natural equilibrium direction concerns the characterization of the KW entanglement asymmetry in Haar-random states. In particular, it would be interesting to understand how the typical value of the asymmetry depends on the symmetry properties of the ensemble, generalizing the recent results obtained for invertible symmetries~\cite{ampc-24,rac-24,rac-25,csm-26,yang-26, gotta-26}.

The nonequilibrium results presented here also raise several open questions. While we found that the Kramers-Wannier symmetry is dynamically restored after quenches to the critical point, it remains unclear how general this phenomenon is. In particular, ordinary invertible symmetries need not be restored after a quench to a symmetric Hamiltonian~\cite{amvc-23}. Whether analogous examples exist for non-invertible symmetries, or whether their algebraic structure imposes stronger constraints on the dynamics, remains an intriguing open problem. More broadly, it would be interesting to understand whether the quasiparticle framework developed here can be extended to interacting integrable models, where it could provide a universal description of the restoration of non-invertible symmetries beyond free theories.

It would also be interesting to understand how the present results extend beyond integrable systems to quantum-chaotic models. In this regard, random quantum circuits provide a particularly attractive framework, as they offer both a generic description of chaotic dynamics and an analytically tractable setting. Such systems have already proved extremely fruitful in the study of entanglement asymmetry for invertible symmetries~\cite{liu-24,tcd-24,yu-25,amcp-25,fcb-24,amt-26, li-26}.

Overall, our work establishes entanglement asymmetry as a powerful probe of non-invertible symmetries both in and out of equilibrium. We hope that it will stimulate further investigations of the interplay between dualities, quantum information, and nonequilibrium dynamics in many-body quantum systems.

\section*{Acknowledgments}
We thank Francesco Benini for very useful discussions.
M. V. gratefully acknowledges the hospitality of SISSA for an internship, during which the present project was conceived.
F.A. and P.C. acknowledge financial support from the European Commission through the ERC-AdG grant MOSE No.\ 101199196.

\appendix

\section{Asymptotic behavior of $\Tr\rho_A\hat{\rho}_A$}
\label{app:trrhohat}

In this appendix, we derive the large-subsystem asymptotics of the mixed overlap
$\Tr\rho_A\hat{\rho}_A$. As discussed in Sec.~\ref{sec4}, this quantity can be expressed in terms of the correlation matrices of the reduced density matrices as
\begin{equation}
\Tr\rho_A\hat{\rho}_A
=
\sqrt{
\det\left(
\frac{I+\Gamma\hat{\Gamma}}{2}
\right)
}.
\label{trrhohat_corr}
\end{equation}
The problem is therefore reduced to determining the asymptotic behavior of the determinant of the matrix $I+\Gamma\hat{\Gamma}$.

Both $\Gamma$ and $\hat{\Gamma}$ are block-Toeplitz matrices with $2\times2$ blocks. We denote a generic block-Toeplitz matrix by
\begin{equation}
(T_\ell[g])_{jj'}
=
\int_0^{2\pi}
\frac{dk}{2\pi}
e^{-ik(j-j')}
g(k),
\qquad
j,j'=1,\ldots,\ell,
\end{equation}
where $g(k)$ is a $d\times d$ matrix-valued symbol, with $d=2$ in the present case.

Strictly speaking, the product of two block-Toeplitz matrices is not itself block-Toeplitz, so the Widom-Szegő theorem~\cite{WIDOM1974284} cannot be directly applied to the matrix $\Gamma\hat{\Gamma}$. Nevertheless, it is natural to conjecture that the leading extensive contribution to the determinant is still given by the Widom-Szegő formula, namely~\cite{amvc-23}
\begin{equation}
\log\det
\left[
I+
\prod_{m=1}^n
T_\ell[g_m]
\right]
=
A\,\ell+o(\ell),
\label{Widom_conj}
\end{equation}
with
\begin{equation}
A
=
\int_0^{2\pi}
\frac{dk}{2\pi}
\log
\det
\left[
I+
\prod_{m=1}^n
g_m(k)
\right].
\label{Widom_symbol}
\end{equation}
Although we are not aware of a rigorous proof of Eq.~\eqref{Widom_conj}, we have verified its validity numerically to high precision in all the cases considered in this work.

Equation~\eqref{trrhohat_corr} can then be written as
\begin{equation}
\log\Tr\rho_A\hat{\rho}_A
=
\frac12
\log
\det(I+\Gamma\hat{\Gamma})
-
\ell\log2.
\end{equation}
The symbols associated with the two correlation matrices are
\begin{equation}
\mathcal{G}(k)
=
\begin{pmatrix}
0 & -ie^{-i\Delta_k}\\
ie^{i\Delta_k} & 0
\end{pmatrix},
\qquad
\hat{\mathcal{G}}(k)
=
\begin{pmatrix}
0 & -ie^{-i\hat\Delta_k}\\
ie^{i\hat\Delta_k} & 0
\end{pmatrix},
\end{equation}
whose product is simply
\begin{equation}
\mathcal{G}(k)\hat{\mathcal{G}}(k)
=
\begin{pmatrix}
e^{-i(\Delta_k-\hat\Delta_k)} & 0\\
0 & e^{i(\Delta_k-\hat\Delta_k)}
\end{pmatrix}.
\end{equation}
Therefore,
\begin{equation}
\det\!\left[I+\mathcal{G}(k)\hat{\mathcal{G}}(k)\right]
=
\left(1+e^{i(\Delta_k-\hat\Delta_k)}\right)
\left(1+e^{-i(\Delta_k-\hat\Delta_k)}\right)
=
2\left[1+\cos(\Delta_k-\hat\Delta_k)\right].
\end{equation}
Substituting this result into Eq.~\eqref{Widom_symbol} yields
\begin{equation}
A
=
\int_0^{2\pi}
\frac{dk}{2\pi}
\log
\left[
2\left(
1+\cos(\Delta_k-\hat\Delta_k)
\right)
\right].
\end{equation}
Finally, we obtain the asymptotic expression
\begin{equation}
\log\Tr\rho_A\hat{\rho}_A
=
-\ell
\int_0^{2\pi}
\frac{dk}{4\pi}
\left[
\log2
-
\log
\left(
1+\cos(\Delta_k-\hat\Delta_k)
\right)
\right]
+o(\ell).
\label{exponential_decrease_trpphat2}
\end{equation}
Equation~\eqref{exponential_decrease_trpphat2} shows that the overlap $\Tr\rho_A\hat{\rho}_A$ decays exponentially with the subsystem size throughout the gapped phases. As discussed in the main text, the decay rate vanishes only at the self-dual point, where the Kramers-Wannier symmetry is exactly restored.

\section{Long-time behavior of Kramers-Wannier asymmetry}
\label{app:LT}

In this appendix, we derive the long-time asymptotics of the second Rényi Kramers-Wannier entanglement asymmetry following a quench to the critical point.

Since we already established that
\[
\lim_{t\to\infty}\Delta S_A^{(2)}(t)=0,
\]
our goal is to determine the leading correction in the limit
$t\to\infty$. Throughout this derivation we employ the quasiparticle approximation introduced in Sec.~\ref{QP}.

Using Eq.~\eqref{DS2} together with the quasiparticle identity
\[
\Tr\rho_A^2(t)=\Tr\hat{\rho}_A^2(t),
\]
the entanglement asymmetry can be written as
\begin{equation}
\Delta S_A^{(2)}(t)
=
\log
\frac{4\Tr\rho_A^2}
{2\Tr\rho_A^2+2\Tr\rho_A\hat{\rho}_A}
=
-\log
\left[
\frac{1+\Tr\rho_A\hat{\rho}_A/\Tr\rho_A^2}{2}
\right].
\end{equation}
At long times,
\[
\Tr\rho_A\hat{\rho}_A
\longrightarrow
\Tr\rho_A^2,
\]
and therefore
\begin{equation}
\Delta S_A^{(2)}(t)
=
-\log
\left[
1+
\frac{\Tr\rho_A\hat{\rho}_A/\Tr\rho_A^2-1}{2}
\right]
\simeq
\frac12
\left(
1-
\frac{\Tr\rho_A\hat{\rho}_A}
{\Tr\rho_A^2}
\right),
\label{DeltaS_expansion}
\end{equation}
where we expanded the logarithm to first order.

The quasiparticle expressions~\eqref{trace_rho2_dyna}~and~\eqref{tr_rhorhohat_dyna} for the two traces give
\begin{align}
\log
\frac{\Tr\rho_A\hat{\rho}_A}
{\Tr\rho_A^2}
=
-\ell
\int_0^{2\pi}
\frac{dk}{2\pi}
\,f(k)
\left[
1-
\min\!\left(
1,\frac{2|v_k|t}{\ell}
\right)
\right],
\label{ratio_traces}
\end{align}
where
\[
f(k)
=
\frac12
\left[
\log2-
\log
\bigl(
1+\cos(\Delta_k-\hat\Delta_k)
\bigr)
\right].
\]
Since the integrand in Eq.~\eqref{ratio_traces} vanishes whenever
$2|v_k|t/\ell>1$, only the slowest quasiparticles contribute to the long-time behavior. The quasiparticle velocity vanishes only at
$k=\pi$, and therefore we expand around this point by setting
\[
u=\pi-k.
\]
Using
\[
v_k
=
\varepsilon_k'
\sim
\frac{|u|}{2},
\]
we find that the relevant integration region is
$|u|\lesssim\ell/t$.

We also need the expansion of $f(k)$ near $k=\pi$.
Using
\[
\cos(\Delta_k-\hat\Delta_k)
=
\frac{2h_0-(1+h_0^2)\cos k}
{1+h_0^2-2h_0\cos k},
\]
one obtains
\begin{equation}
\cos(\Delta_k-\hat\Delta_k)
=
1-
\frac{u^2}{2}
\left(
\frac{h_0-1}{h_0+1}
\right)^2
+O(u^4),
\end{equation}
which implies
\begin{equation}
f(k)
=
\frac{u^2}{8}
\left(
\frac{h_0-1}{h_0+1}
\right)^2
+O(u^4).
\end{equation}

Substituting these expansions into Eq.~\eqref{ratio_traces}, using the evenness of the integrand, and introducing the scaling variable
\[
u=\frac{\ell}{t}x,
\]
we obtain
\begin{equation}
\Delta S_A^{(2)}(t)
\simeq
\frac{\ell^2}{8\pi t}
\left(
\frac{h_0-1}{h_0+1}
\right)^2
\int_0^1dx\,
x^2(1-x)
\left(
\frac{\ell}{t}
\right)^2
=
\frac{\ell^4}{16\pi t^3}
\left(
\frac{h_0-1}{h_0+1}
\right)^2
\int_0^1dx\,x^2(1-x).
\end{equation}
Finally,
\[
\int_0^1dx\,x^2(1-x)=\frac1{12},
\]
yielding the asymptotic decay
\begin{equation}
\Delta S_A^{(2)}(t)
\underset{t\to\infty}{\sim}
\frac{\ell^4}{192\pi t^3}
\left(
\frac{h_0-1}{h_0+1}
\right)^2,
\end{equation}
which is the result reported in the main text.

\end{document}